\documentclass[apj,numberedappendix]{emulateapj}

\newcommand{\be}{\begin{eqnarray}}
\newcommand{\ee}{\end{eqnarray}}
\newcommand{\lp}{\left(}
\newcommand{\rp}{\right)}

\newcommand{\E}[1]{\times10^{#1}}
\newcommand{\smpy}{ \ M_\odot \ {\rm yr}^{-1}}
\newcommand{\mdot}{\dot{M}}
\newcommand{\porb}{P_{\rm orb}}
\newcommand{\msol}{ \ M_\odot}
\newcommand{\lsol}{ \ L_\odot}

\begin{document}


\shorttitle{He WDs IN CVs}
\shortauthors{SHEN, IDAN, \& BILDSTEN}


\title{Helium Core White Dwarfs in Cataclysmic Variables} 
\author{Ken J. Shen\altaffilmark{1}, Irit Idan\altaffilmark{2,3}, and Lars
  Bildsten\altaffilmark{1,4}}
\altaffiltext{1}
{Department of Physics, Broida Hall, University of California, Santa
  Barbara, CA 93106}
\altaffiltext{2}{Rafael, 31021 Haifa, Israel}
\altaffiltext{3}{Department of Physics, Technion-Israel Institute of Technology, 32000 Haifa, Israel}
\altaffiltext{4}
{Kavli Institute for Theoretical Physics, Kohn Hall, University of California, Santa
  Barbara, CA 93106}


\begin{abstract}

Binary evolution predicts a population of helium core ($M<0.5 \msol$) white dwarfs (WDs) that are slowly accreting hydrogen-rich material from low mass main sequence or brown dwarf donors with orbital periods less than four hours. Four binaries are presently known in the Milky Way that will reach such a mass-transferring state in a few Gyr. Despite these predictions and observations of progenitor binaries, there are still no secure cases of helium core WDs among the mass-transferring cataclysmic variables (CVs).  This led us to calculate the fate of He WDs once accretion begins at a rate $\dot M<10^{-10} \smpy$ set by angular momentum losses.  We show here that the cold  He core temperatures ($T_c<10^7 \ {\rm K}$) and low $\dot M$ result in $ \sim 10^{-3} \msol$ of accumulated H-rich material at the onset of the thermonuclear runaway. Shara and collaborators noted that these large accumulated masses may lead to exceptionally long classical nova (CN) events.  For a typical donor star of $0.2 \msol$, such binaries will only yield a few hundred CNe, making these events rare amongst all CNe. We calculate the reheating of the accreting WD, allowing a comparison to the measured WD effective temperatures in quiescent dwarf novae and raising the possibility that WD seismology may be the best way to confirm the presence of a He WD.  We also find that a very long ($>1000$ yr) stable burning phase occurs after the CN outburst, potentially explaining enigmatic short orbital period supersoft sources like RX J0537-7034 ($\porb=3.5$ hr) and 1E 0035.4-7230 ($\porb=4.1$ hr). 

\end{abstract}



\section{Introduction} 
\label{sec:hewds}

For short enough orbital periods, the evolution of a low-mass star in a binary system up
the red giant branch can be halted by the onset of catastrophic
filling of its Roche lobe. This exposes the low-mass ($M<0.5 \msol $; e.g., \citealt{dcls99,scw02,sf05}) helium core of the red giant and triggers a common envelope
event (see \citealt{sand2000} for an overview) that brings the resulting
cooling He white dwarf (WD) much closer to its stellar companion
\citep{dek92,dekr93,ibtut93}. The discovery by \cite{ma95} that
low-mass WDs have stellar companions supports this origin for many
He core WDs. 

Angular momentum losses (either from gravitational waves or stellar wind
braking) after the common envelope event will drive the binary closer,
with some reaching contact in a few Gyr to become cataclysmic variables (CVs). \cite{sch03} studied such pre-CVs, noting that a few have He WDs.  The recently discovered
system WD 0137-349 is a new pre-CV example with a $0.39 \msol$ WD
and a $0.053 \msol$ brown dwarf companion that will come into contact
in 1.4 Gyr \citep{maxted06}. Population synthesis calculations
\citep{dek92, pol96,hnr01} predict that as many as $\approx 20\%$ of
CVs with orbital periods $\porb <2 \ {\rm hr}$ should have He WDs. However, where
WD masses are inferred in CVs \citep{pa01,tg09}, no such low masses
have been identified.

These considerations led \cite{shara93} to examine classical nova
(CN) events in these unusual systems. They calculated the outcomes for
a $0.4 \msol$ He WD accreting at a rate of $\dot M=10^{-9} \smpy $, finding accumulated masses of $M_{\rm ign}= 9\times
10^{-4} \msol$ for a WD core temperature of $T_c=10^7 \ {\rm K}$ and a
time between outbursts of  $10^6$ yr. They evolved the WD for
nine flashes, finding slow novae followed by a prolonged bright
($L>10^3 \lsol $) phase after the outburst, as the residual $\approx
10^{-4} \msol$ hydrogen shell 
burns stably. The combination of a
large $M_{\rm ign}$ in a tight binary may make these events
exceptional. In the case of C/O WDs, \cite{ibtut92} suggested that
such events would trigger a common envelope event of some
significance, potentially explaining the very bright ``red variable''
seen in M31 \citep{rich89,bosmun04}.

The known pre-CVs with He WDs provide excellent starting points for
our work, so we begin in \S \ref{sec:masstransfer} by noting the
stability of mass transfer once contact is reached and deriving the
subsequent $\dot M(t)$ histories. We find that lower accretion rates
than studied by \cite{shara93} need to be considered, so in \S
\ref{sec:example} we perform (and analyze) a few Gyr of time
evolution at constant low $\mdot = 10^{-10} \smpy$ and
$10^{-11} \smpy$ on a $0.4 \msol$ He WD. Given the low $\dot
M$, there is time for diffusion to occur between
the fresh material and the WD \citep{paq86,ifm92}, which we discuss in the Appendix.

In \S \ref{sec:fullscenario}, we perform calculations of two
specific scenarios motivated by the known He WD pre-CVs. Our focus
there is on the long-term evolution of their CN outbursts as well as the
$T_c$ evolution. We also discuss
the WD surface temperatures when in quiescence as
dwarf nova systems and the potential discovery of He WDs through their
pulsations \citep{arras06}.  In \S \ref{sec:indiv}, we explore details of the individual novae, describing the physics of the convective burning phase in \S \ref{sec:conv}, and the composition of the nova ejecta in \S \ref{sec:comp}, which is not enhanced in CNO nuclei due to the lack of a reservoir of underlying C/O as in normal CNe \citep{stsk72,gehrz98}.  The CN event leaves a remnant burning envelope on the surface of the WD, allowing for a prolonged stable burning phase \citep{shara93} that we detail in \S \ref{sec:supersoft}, which may
explain the few $\porb  \lesssim 4 $ hr long-lived supersoft
sources (SSSs). We conclude in \S \ref{sec:conclusions} by highlighting
where future CV and CN observations may reveal the long-lost helium
core WDs in mass transferring binaries. 


\section{Stable Mass Transfer Rates} 
\label{sec:masstransfer}

The common envelope event that exposes the He WD leaves a wide range
of WD and donor masses ($M_d$), only a fraction of which will
become stable mass-transferring binaries once the donor comes into
contact \citep{dek92, pol96,hnr01}. Table \ref{tab:preCVs} shows the parameters of
those pre-CVs with probable He WDs whose companions will overflow their Roche lobes in $t_{\rm in}< 20$
Gyr, under the assumption of angular momentum losses due to gravitational waves. The prevalence of low-mass ($M_d<0.25 \msol$) main sequence
companions is likely due to their slower inspiral under gravitational
wave losses compared to those with more massive ($M_d>0.25 \msol$)
companions that would have stellar wind braking \citep{sch03}. As we will show in the following sections, all binaries in Table \ref{tab:preCVs} will likely transfer matter stably after contact,
leading to the accretion of cosmic-mix material onto a He WD at $\mdot = 1-4 \E{-11} \smpy$.


\subsection{Stability of Mass Transfer} 

When the main sequence star comes into contact, the
possibility exists for dynamically unstable mass transfer, which
occurs when the donor's Roche radius shrinks faster or expands slower
than the donor's radial response to adiabatic changes \citep{web85,hw87}.
Low-mass main sequence stars have small radiative cores and large
convective envelopes that make up $>50\%$ of the star's mass, and
stars with $M_d<0.3 \msol$ are nearly fully convective.  The
adiabatic response of the radius of a fully convective star to mass
loss is $ \left. d \ln R_d / d \ln M_d \right|_s=-1/3$, and using the
\cite{pac67} approximation to the donor's Roche radius yields $ d \ln
R_{L} / d \ln M_d = -5/3+2q$, where $q \equiv M_d/M$.  Thus,
dynamically unstable mass transfer occurs if the donor is fully
convective and has a mass ratio $M_d/M > 2/3$.  For the maximum mass He WD with $M = 0.5 \msol$, this restricts the main sequence donor mass to 
$M_d< 0.33 \msol$ in order to avoid dynamical mass transfer. If the more
exact Roche radius formula from \cite{egg83} is used, the limit changes only
slightly to $M_d/M< 0.63$. Binaries just at the limit of
stability would have donors that fill their Roche lobes as main sequence stars ($M_d=0.33 \msol,
R_d=0.33 \ R_\odot$; \citealt{ribas08}) with $ \porb = 3$
hr. Any bloating of such a donor  could drive $\porb$
slightly longer, potentially to the $3.5-4$ hr seen in a few supersoft sources, as we discuss in \S \ref{sec:supersoft}.

All the pre-CVs in Table \ref{tab:preCVs} have $M_d/M \lesssim 0.63$ and will become stable
mass transfer systems when they come into contact a time $t_{\rm in}$ from now. We derive their mass transfer rates in \S \ref{sec:cvmdot}. There are certainly He WD post-common envelope systems with less extreme mass ratios that would trigger a dynamical mass transfer event at the onset of Roche lobe filling. These yield a red giant configuration, as the supply of freshly accreted H leads to a rejuvenation of the H burning shell typical of a He core on the first ascent of the red giant branch. 

\begin{table*}
	\begin{center}
	\caption{Helium WD Pre-CVs with $t_{\rm in}<20$ Gyr (as of July 2009)}
	\begin{tabular}{|c|c|c|c|c|}
	\hline
	System & $M$ & $M_d$ &  $t_{\rm in} $ & Ref.\\
	 & ($M_\odot$) &  ($M_\odot$) & (Gyr) & \\
	\hline
	\hline
	SDSS J1435+3733 & $0.48-0.53$ & $0.19-0.25$ & $0.75-1.1$ &1,2 \\
	\hline
	HR Cam (GD 448)  & $0.41 \pm 0.01$ & $0.096 \pm 0.004 $ & $1.5-2.1$ &3\\
	\hline
	WD 0137-349 & $0.39 \pm 0.035$ & $0.053 \pm 0.006 $ & $1.7-2.5$ & 4 \\
	\hline
	SDSS J1529+0020 & $0.40 \pm 0.04$ & $0.25 \pm 0.12$ & $0.54-6.3$& 5\\
	\hline
	SDSS J0110+1326 & $0.47 \pm 0.2$ & $0.255-0.38$ & $12-19$ & 2 \\
	\hline
	SDSS J1724+5620 & $0.42 \pm 0.01$ & $0.25 -0.38$ & $13-20$ & 5\\
	\hline
	RR Cae & $0.44\pm0.022$ & $0.182\pm0.013$ & $18-23$ & 6 \\
	\hline
	SDSS J1212-0123 & $0.33-0.48$ & $0.26 \pm 0.03 $ & $16-26$ & 7 \\
	\hline
	\end{tabular}
	\label{tab:preCVs}
	\end{center}
1. \cite{stein08},
2. \cite{pyr09},
3. \cite{maxted98},
4. \cite{maxted06},
5. \cite{rebassa08},
6. \cite{maxted07},
7. \cite{nebo09}
\end{table*}


\subsection{ Mass Transfer Rates for Stable Scenarios}
\label{sec:cvmdot}

Our focus here is on those systems that undergo stable mass transfer,
with the WD accumulating mass over time and then ejecting most of it
during the CN. We start with a total mass
$M_t=M_d+M$ and orbital angular momentum $J=M_dM(Ga/M_t)^{1/2}$ for an
orbital separation $a$.  Mass transfer is driven by
the rate of orbital angular momentum loss, $\dot J$, via 
\begin{equation}
{\dot J\over J}={\dot M_d\over M_d}+{\dot M\over M} + \frac{1}{2} {\dot a\over
  a}- \frac{1}{2} {\dot M_t\over M_t}.
\end{equation} 
Since CNe lead to ejection of material from the binary, we must account for mass loss. We define an excavation factor $f$ by
setting $\dot M_t=f\dot M_d$, so that $f=1$ means the WD, on average,
keeps a constant mass (i.e., the CNe eject the amount of matter that
is accreted), and $f=0$ means that the WD keeps all the accreted mass. The
resulting relation is then 
\begin{equation}
{\dot J\over J}={\dot M_d\over M_d}\left[1+{(f-1)M_d\over
    M}-\frac{f}{2} {M_d\over M_t}\right]+ \frac{1}{2} {\dot a\over a}.
\end{equation} 
Presuming that the donor always fills its Roche lobe, and using the 
\cite{pac67} formulation, we obtain
\begin{equation}
{\dot J\over J}= \frac{1}{2} {\dot R_d\over R_d}+
{\dot M_d\over M_d}\left[{5\over 6}+{(f-1)M_d\over
    M}- \frac{f}{3} {M_d\over M_t}\right].
\end{equation}
We need the response of the donor star's radius to mass loss, $\zeta_d=
d \ln R_d / d \ln M_d $, which initially occurs on a timescale longer
than the Kelvin-Helmholtz time, allowing the star to stay close to the
main sequence (e.g., $\zeta_d\approx 1$). However, the increase of the
Kelvin-Helmholtz time as $M_d$ decreases leads to a nearly adiabatic
response ($\zeta_d\approx -1/3$) late in the binary evolution
\citep{kolbbar99}.

\begin{figure}
	\plotone{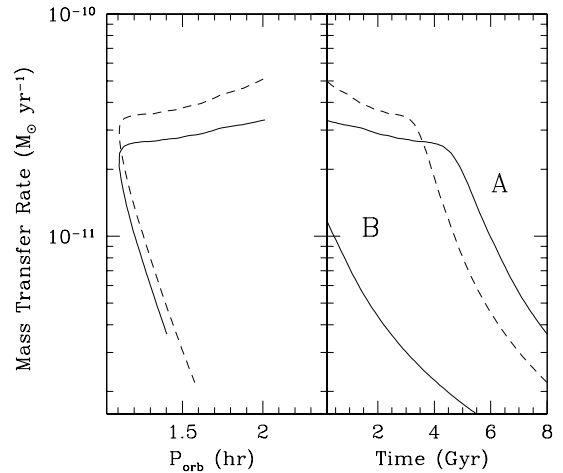}
\caption{The solid line in the left panel shows $\dot M$ as a function
  of $\porb $ assuming that all mass accreted onto the
  $0.4 \msol$ WD is ejected by CNe ($f=1$) for an
  initial $M_d=0.2 \msol$ donor (scenario A; e.g., SDSS J1529+0020). The dashed line assumes that
  all accreted matter stays on the WD ($f=0$).  The upper solid and dashed
  lines in the right-hand panel show the corresponding $\dot M(t)$
  profiles.  The lower solid curve in the right hand panel shows $\dot
  M$ for a donor with initial $M_d=0.05 \msol$ (e.g., WD 0137-349 ),
  assuming that CNe eject all the accreted material ($f=1$; scenario B).}
	\label{fig:mdotplot}
\end{figure}

The resulting mass transfer rate is then
\begin{equation}
{\dot J\over J}={\dot M_d\over M_d}\left[{5\over 6}+{\zeta_d\over 2}+{(f-1)M_d\over
   M}- \frac{f}{3} {M_d\over M_t}\right],
\end{equation}
where we use the work of \cite{kolbbar99} to obtain $R_d(M_d)$ and $\zeta_d$.  Their $R_d(M_d)$ relation was calculated self-consistently for a $M=0.6 \msol$ WD with gravitational wave losses alone.  We also presume that gravitational-wave losses set $\dot J$ and consider a few specific scenarios in Figure \ref{fig:mdotplot}.  The two solid lines labeled A and B are the two scenarios we will discuss at length in \S \ref{sec:fullscenario}. Scenario A begins with a $M_d=0.2\msol$ donor, while scenario B begins with a $M_d=0.05 \msol$ donor.  Both assume that all matter accreted onto the WD is ejected during the CN event.  The resulting accretion rates are $< 4\E{-11} \smpy$ once mass transfer begins.  The trends that we discuss throughout the rest of the paper for $0.2 \msol$ donors will apply to larger donors as well because the accretion rates determined by gravitational wave radiation for low-mass main-sequence donors differ by a factor of a few at most.  For example, the accretion rate for the most massive stably mass-transferring system ($M_d = 0.33 \msol$, $M = 0.5 \msol$) is only 3 times larger at a given orbital period than our fiducial example with $M_d = 0.2 \msol$ and $M=0.4 \msol$.

We neglect any potential variations of $\dot M$ on the CN recurrence time due to prolonged periods of ``hibernation'' \citep{shara86}.  We further note that the $R_d(M_d)$ relation we use assumes a donor with solar metallicity.  As \cite{skr97} and \cite{kolbbar99} note, the secondary's composition does have an effect on its structure and thus on the secular evolution of the CV.  We have also neglected the small correction needed to re-adjust the $R_d(M_d)$ relation for our specific $\dot M$ history, which is lower by a factor of $\approx 1.5$ than that in \cite{kolbbar99} due to our $0.4 \msol$ WD (they modeled a $0.6 \msol$ WD).  This would tend to slightly shorten the minimum orbital period, as the mass-transfer timescale at a given donor mass is longer in our case.  However, given that the current calculations do not match the now observed minimum orbital period in either case \citep{gaen09}, we will neglect the corrections due to the secondary's composition and $\dot{M}$ history.

\section{Reheating of slowly accreting helium white dwarfs}
\label{sec:example} 

The previous section showed that the relevant accretion rates are much lower than that considered by \cite{shara93}. In addition, by the time the known pre-CVs come into contact and initiate stable mass transfer, the He WD core will have cooled to $T_c\approx 2-6\times 10^6\ {\rm K}$ \citep{altben97}. Extensions of the work by \cite{tb04} on CN ignition masses to lower $M$, $\dot M$ and $T_c$ yields $M_{\rm ign}\approx 10^{-3} \msol$. This implies CN recurrence times of $10^7-10^8$ yr during the many gigayears of evolution under accretion.  The long evolution time allows for significant thermal coupling between the freshly accreted H envelope and the He WD core during the accumulation of fresh matter. For this reason, we must track $T_c$ and its evolution under the action of accretion for a prolonged period of hundreds of CN events.


\subsection{Analytic estimate of thermal coupling between the envelope and core}

\label{sec:thermev}

Due to the long accumulation and evolution time, much of the He WD core participates in the thermal balance of the CN cycle. We begin by estimating the timescale for heat transport between two mass shells at radial coordinates $r_0$ and $r$ \citep{hl69}, 
\be
	\tau_d \approx  \frac{3}{16 a_{\rm SB}c} \left[ \int_{r_0}^r \lp \frac{\kappa c_P }{T^3} \rp^{1/2} \rho dr \right]^2 , 
	\label{eq:difftime}
\ee
where $c_P$ is the specific heat, $a_{\rm SB}$ is the Stefan-Boltzmann
radiation constant, and $\kappa$ is the opacity.
Electron conduction determines heat transfer in the degenerate He
core, so $\kappa = \kappa' T^2/\rho^2$, where $\kappa' \approx 6\E{-7}
{\rm \ g \ cm^{-4} \ K^{-2} }$ from fitting to \cite{cass07}'s opacities at
$\rho = 2\E{5}$ g cm$^{-3}$ and $T=6\E{6}$ K. When the 
core is isothermal with a constant $c_P$, the thermal
diffusion time becomes
\be
	\tau_d &\approx& \frac{3\kappa' c_P}{16a_{\rm SB}cT_c} (r-r_0)^2 \nonumber \\
	&\approx & 1.6\E{8} {\rm \ yr \ } \lp \frac{\kappa'}{6\E{-7} {\rm \ g \ cm^{-4} \ K^{-2} } } \rp \nonumber \\
	&& \times \lp \frac{6\E{6} {\rm \ K}}{T_c} \rp \lp  \frac{c_P}{3k_B/4m_p} \rp \lp \frac{r-r_0 }{10^9 {\rm \ cm}} \rp^2 ,
	\label{eq:difftimeest}
\ee
where we have normalized $c_P$ to the dense liquid value for pure
helium (i.e., an internal energy of $3k_BT$ per ion, where $k_B$ is
Boltzmann's constant and $m_p$ is the baryon mass).\footnote{For
low-mass He WDs, the Coulomb coupling parameter $\Gamma$ is
$\approx 10$, so the ions are in a liquid state, but not necessarily deep
into the $3k_BT$ limit.  Thus, the actual values for $c_P$ are
 between the ideal gas and liquid values.}

\begin{figure}
	\plotone{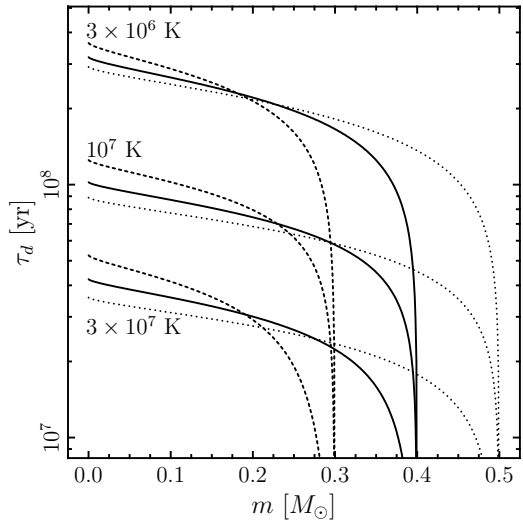}
	\caption{Thermal diffusion time from the outside of the core
          to interior mass points for $M=0.3 \msol $ (dashed lines), $0.4 \msol$
          (solid lines), and $ 0.5 \msol$ (dotted lines)
           He WDs. The isothermal core temperatures are marked.}
	\label{fig:difftime}
\end{figure}

We numerically integrate equation (\ref{eq:difftime}) with the full
dependences of $\kappa$ and $c_P$ for varying masses
and isothermal temperature to find the thermal diffusion times from the
core-envelope boundary to a given mass coordinate within the WD. These are shown in Figure \ref{fig:difftime} for $0.3 \msol$ (dashed lines), $0.4 \msol$ (solid lines), and $0.5 \msol$
(dotted lines) He WDs.  Each set of lines shows different
core temperatures.  From top to bottom, these are $3\E{6}$, $10^7$,
and $3\E{7}$ K. The only dependence on $M$ in equation
(\ref{eq:difftimeest}) is from the radius of the core-envelope interface, which are all within
$10-20$\% of each other for these low-mass WDs. Our simple estimate in equation (\ref{eq:difftimeest}) is fairly close to the integrated results in Figure \ref{fig:difftime}; a core temperature of $6\E{6}$ K yields $\tau_d \approx 2\E{8}$ yr, and raising $T_c$ by a factor of 10 decreases the thermal diffusion time by almost a
factor of 10, as predicted by equation (\ref{eq:difftimeest}).


\subsection{Numerical Evolution of the Thermal State for Constant $\dot{M}$}
\label{sec:constMdot}

In this section, we describe several time-dependent calculations of constant $\dot{M}$ onto a $0.4 \msol $ He WD in order to discuss the thermal evolution and compare to the work of \cite{tb04}.  The specific outcomes for the $\dot M$ histories appropriate to CVs calculated in \S \ref{sec:cvmdot} will be discussed in \S \ref{sec:fullscenario}.

Numerical calculations were performed with the
hydrodynamic Lagrangian stellar evolution code of \cite{prialkov95},
which contains an extended nuclear reaction network of 40 elements up
to $^{31}$P \citep{cf88,angu99,tim99} and implements the OPAL opacities \citep{ir93,ir96}. Convection is treated by mixing length theory with a mixing-length parameter $\alpha=2$ \citep{miha78,mazz79,vand83}.  Inter-species diffusion is allowed to
occur (see the Appendix for discussions of diffusion), but no other mixing processes are presumed.  No $^3$He is included in the newly accreted material \citep{shar80,tb04,sb09a}, but its presence in the incoming material should be negligible as the donor is unevolved.  The CN
mass-loss phase which follows the phase of rapid expansion uses a steady optically-thick supersonic wind solution \citep{prial86}.  Common envelope effects are neglected.

The $0.4 \msol$ He WD has a core composition of $98\%$ He and $2\%$ $^{14}$N by mass and is surrounded with an initially thin envelope of solar composition.  Three cases were studied, with different $\mdot$ and initial $T_c$: $\mdot=10^{-10} $ and $10^{-11} \smpy$ for initial $T_c = 6\E{6}$ K, and $\mdot=10^{-11} \smpy$ for initial $T_c = 8\E{6}$ K.  We numerically follow the evolution through hundreds of nova outbursts, with each nova cycle calculated through all evolutionary phases. \cite{yaron05} found that novae at these low accretion rates tend to erode the WD ($f>1$). Therefore, after each outburst and resulting envelope ejection, the outer mass shell of the stripped WD core is split into multiple shells of $\lesssim 10^{-7} \msol$ so that the next nova cycle can be accurately calculated.

 Equation (\ref{eq:difftimeest}) and Figure \ref{fig:difftime}
 indicate that the timescale for thermal coupling between the core and
 envelope is of order the timescale between nova outbursts for
 low $\dot{M}=10^{-11} \smpy$, although not for the higher
 $\dot{M}=10^{-10} \smpy$ case, which has a shorter accumulation time of $\approx
 6\E{6}$ yr.  However, because the envelope has a much
 lower mass (and thus a smaller thermal content) than the core, it
 will take many nova cycles to significantly increase $T_c$. The
 thermal content in the nearly ideal gas envelope of mass $M_{\rm
   env}$ with mean atomic weight $\mu\approx 0.6 $ is $E_{\rm
   th,env}=3 M_{\rm env} k_BT/2\mu m_p\approx 4\times 10^{45} {\rm
   \ erg} \ (T/10^7 \ {\rm K})(M_{\rm env}/10^{-3} \msol)$, whereas the thermal
 content in the He core is $E_{\rm th,core}\approx 3 Mk_BT_c/4
 m_p\approx 5\times 10^{47} {\rm \ erg} \ (T_c/10^7 {\rm \ K})(M/0.4 \msol)$. The
 100-fold ratio of these thermal energies implies that it will take
 $\sim 100$ nova cycles to significantly heat the core if the only energy source is thermal.  Some nuclear ``simmering'' could shorten the required time, as we discuss later in this section.

\begin{figure}
       \plotone{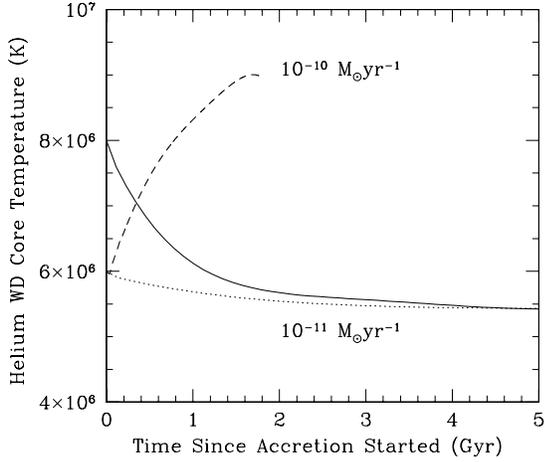}
       \caption{Time evolution of the core temperature, $T_c$,  of the $0.4 \msol$
   He WD for two different constant accretion rates and initial
   core temperatures. The solid (dotted) lines shows the
$\dot M=10^{-11} \smpy$ case with initial core temperatures of
   $8\times 10^6 \ {\rm K}$ ($6\times 10^6 \ {\rm K}$). The dashed
   line is for an initial temperature of $6\times 10^6 \ {\rm K}$,
   but $\dot M=10^{-10} \smpy$.  Lines are smoothed on a timescale of several nova cycles in order to make the secular evolution more clear.}
       \label{fig:tcoreevol}
\end{figure}

Figure \ref{fig:tcoreevol} shows the time evolution of $T_c$ for the three cases, all eventually reaching a final equilibrium core temperature.  The core temperatures have been smoothed in order to make the evolution more clear.  The dashed line is for $\dot M=10^{-10}\smpy$, which reaches an equilibrium temperature of $T_c\approx 9\times 10^6$ K and surface luminosity in disk quiescence $L_q \approx 7\E{-3} \lsol $.  The time needed to reach this equilibrium is about that expected from the earlier considerations regarding the thermal content and heating of the core from the accumulated envelope, as $\approx 0.2 \msol$ of matter was put on the WD through this evolution, triggering 200 CNe.

\begin{figure}
	\plotone{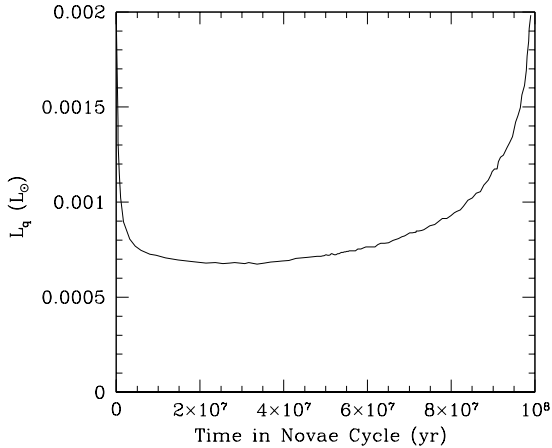}
	\caption{Time evolution of the quiescent surface luminosity, $L_q$, for a $0.4 \msol $ He WD accreting at $10^{-11} \smpy$ after it has reached its equilibrium core temperature.}
	\label{fig:lqui}
\end{figure}

The evolutions for lower $\dot M=10^{-11} \smpy$ are shown by the solid and dotted lines in Figure \ref{fig:tcoreevol} for two different initial core temperatures.  For these cases, the initial $T_c$ is higher than the equilibrium value; accretion serves to slow the cooling of the WD as opposed to heating the core in the higher $\mdot = 10^{-10} \smpy$ case.  The equilibrium core temperature is $T_c=5.5\E{6}$ K with a quiescent surface luminosity of $L_q\approx 7\E{-4} \lsol$ for the majority of the accretion phase of each nova cycle, as shown in Figure \ref{fig:lqui}, which plots the evolution of the quiescent surface luminosity during a typical nova cycle after the equilibrium core temperature has been reached.

\cite{tb04} and \cite{eykp07} showed that equilibrium core temperatures would be reached under the action of accretion and CN events but did not consider WD masses this low. \cite{tb04} showed that the equilibrium core temperatures could be understood as that temperature where the cooling of the core (mostly during the accumulation stage) would be matched by the heating of the core (typically late in the accumulation, approaching the CN event). They highlighted two energy sources during accumulation: ``compressional'' heating (really entropy loss as material is advecting into the star) and slow nuclear burning.  Following the derivation in Appendices A and B of \cite{tb04}, which in turn follows \cite{ns77}, \cite{nomo82a}, and \cite{hern88}, yields the equivalent of \cite{tb04}'s equation (B4) for the ``compressional'' heating luminosity,
\be
	\label{eq:lcomp}
	L_q &\approx& 4.5{k_BT_c \over m_p}\dot M \nonumber \\
	&\approx& 6.1\E{-4} \lsol \lp T_c\over 10^7 {\rm \ K} \rp \lp \frac{\mdot}{10^{-11} \smpy} \rp ,
\ee
after accounting for the differences in the Coulomb coupling parameter, mean atomic weight, and temperature in the core, and the larger $M_{\rm ign}$.  This predicts $L_{\rm comp} = 3.4\E{-4} \lsol $ and $L_{\rm comp} = 5.5\E{-3} \lsol$ for $\dot M=10^{-11}\smpy$ and $\dot M=10^{-10}\smpy$, respectively, after their equilibrium core temperatures have been reached.  These estimates for the outgoing luminosity are $\approx 50\%$ lower than Figure 4 shows for $\mdot = 10^{-11} \smpy$ and $\approx 20\%$ lower than the numerically calculated $L_q=7\E{-3} \lsol $ for $\mdot = 10^{-10} \smpy$.

These differences in luminosity can be attributed to our numerical calculation's inclusion of nuclear simmering.  \cite{tb04} showed the potential of nuclear energy release during accumulation especially for massive envelopes like these (see their Figure 3). However they did not carry out this specific calculation.  Our time dependent calculations allow us to see how the nuclear energy source can come into play.

\begin{figure}
       \plotone{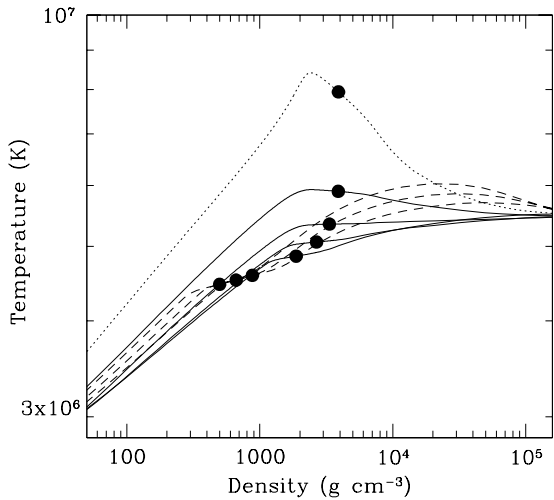}
       \caption{Density and temperature evolution during the period
         after a CN and the subsequent accumulation to the next
         explosion for a $0.4 \msol$ WD accreting at $\dot
         M=10^{-11}\smpy$. The three dashed lines are for times $5\times
         10^6 {\rm \ yr}$, $7\times 10^6 {\rm \ yr}$, and $10^7 {\rm \ yr}$
         after the CN, from top to bottom. The solid lines, from bottom to top, are for times $3\times 10^7 {\rm \ yr}$,
         $5\times 10^7 {\rm \ yr}$, $7\times 10^7 {\rm \ yr}$, and
         $9\times 10^7 {\rm \ yr}$ after the last outburst. The dotted
         line is the profile at a time of $9.95\times 10^7 {\rm \ yr}$,
         when convection has started.  Bullets show the base of the accreted envelope.}
       \label{fig:rhozone}
\end{figure}

Figure \ref{fig:rhozone} shows snapshots of the thermal
profile during different stages of the $10^8$ yr CN cycle for $\mdot = 10^{-11} \smpy$ after the WD has reached its equilibrium core temperature. The three
dashed lines (top to bottom) are for times $5\times 10^6 {\rm \ yr}$,
$7\times 10^6 {\rm \ yr}$ and $10^7 {\rm \ yr}$, after the CN. The
temperature inversion shows that heat flows from the hot post-CN envelope into the region beneath the burning layer during this time, to exit later during the first part of the accumulation phase. After this heat has been radiated away, the thermal profiles
(three lower solid lines at $3\times 10^7 {\rm \ yr}$, $5\times 10^7
{\rm \ yr}$ and $7\times 10^7 {\rm \ yr}$ after the outburst) begin to resemble cooling WD
profiles.  Accumulation of material (and nuclear simmering) redevelops
the temperature inversion (top solid line at $9\times 10^7 {\rm \ yr}$),
and the dotted line shows the profile at $9.95\times 10^7 {\rm \ yr}$,
when convection has initiated.  Simply put, the excursions of the
temperature during the CN event and the long accumulation time
allow for some nuclear energy to be transported into the outer layers of the WD core over
time, to be radiated later in the accumulation stage.

\section{Thermal Evolution for CV Scenarios and Dwarf Novae in Quiescence}
\label{sec:fullscenario}

As discussed in \S \ref{sec:masstransfer}, stable mass transfer for low mass He WDs requires a low-mass companion that will transfer mass at the rate set by gravitational radiation losses. The more rapid mass transfer systems (i.e., above the period gap) that have stellar wind braking will be rarer. Our focus in this section is on the evolution of the two scenarios discussed in \S \ref{sec:masstransfer}, building on the intuition gained in \S \ref{sec:example}.


\subsection{Thermal evolution}
\label{sec:thermevol}

\begin{figure}
	\plotone{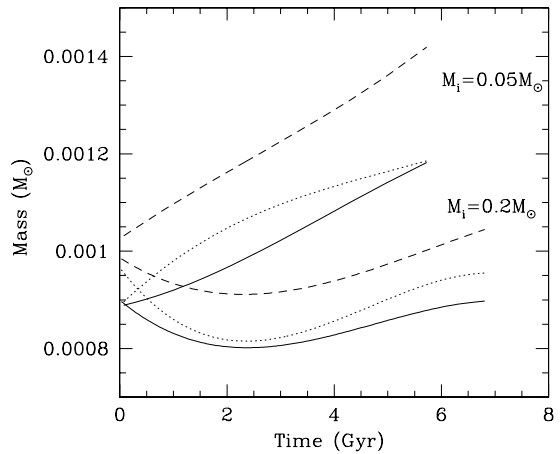}
	\caption{Accreted mass (solid lines), envelope mass above the position where convection first begins (dotted lines), and ejected mass (dashed lines) for the two binary evolution cases of Figure \ref{fig:mdotplot}. Due to the diffusion of H into the He during accumulation, the initial convective envelope mass and ejected mass are always slightly larger than the amount of mass accreted since the last outburst.}
	\label{fig:ejectevol}
\end{figure}

Figure \ref{fig:ejectevol} shows the mass accreted between nova outbursts (solid lines), envelope mass above the position where convection first begins during the thermonuclear runaway (dotted lines), and the ejected mass (dashed lines) for scenarios A and B of Figure \ref{fig:mdotplot}. Even though the mass transfer rates  and core temperatures differ in these scenarios, the typical CN ignition masses change very little. Most importantly, diffusion of the H into the He and the resulting convection during the outburst lead to an ejected mass that is slightly larger than what was accreted. Hence, our earlier approximation that the accreted mass is ejected was appropriate.

\begin{figure}
       \plotone{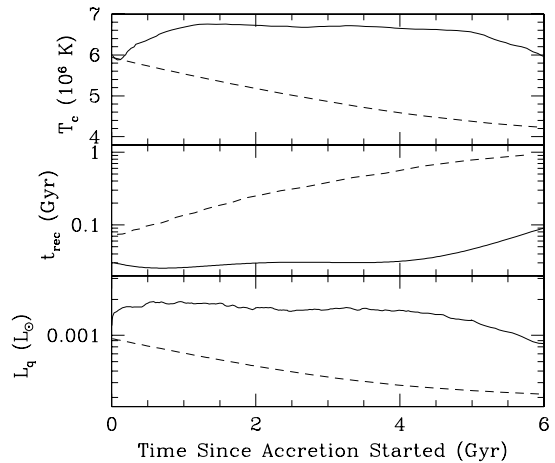}
       \caption{The top panel shows the mass-averaged WD core temperature as a function of
         time for a $0.4 \msol$ He WD with initial $T_c = 6\times
         10^6 $ K for the two $\dot M$ histories (scenarios A and B) of Figure
         \ref{fig:mdotplot}. The solid (dashed) line is for
         initial $M_d=0.2$ $(0.05) \msol$ donor stars. The middle panel shows the
         time between CN outbursts for the same
         evolutionary scenarios.  The bottom panel shows the evolution of the WD's surface luminosity in disk quiescence, $L_q$. For this plot, we use the minimum of $L_q$ during each nova cycle, which because of the broad minimum is indicative of the
`typical' value (see Figure \ref{fig:lqui}).}
       \label{fig:Tcevol}
\end{figure}

The thermal evolution of the WD core is more complex due to the changing $\dot M$ in these realistic scenarios, as shown in Figure \ref{fig:Tcevol}. The top panel shows the $T_c$ evolution for which the initial donor is a $0.2 \msol$ main sequence star (scenario A; solid line), which exhibits an initial heating to a near-equilibrium core temperature of $T_c=6.8\times 10^6$ K and a luminosity of $L_q \approx 2\times 10^{-3} \lsol$ (solid line in bottom panel) during the 4 Gyr when the accretion rate is nearly constant at $\dot M=3\times 10^{-11} \smpy$; this value of $L_q$ is in agreement with the prediction of Equation (\ref{eq:lcomp}).  This case is intermediate to the two calculations with constant $\dot M$ discussed in \S \ref{sec:thermev}. During these 4 Gyr, the CN recurrence time (solid line in middle panel) is $3\times 10^7$ years for about 130 CN events. After 4 Gyr, the donor begins to expand under further mass loss, and $\dot M$ declines as the binary passes through the period minimum. The WD then cools, and the recurrence time becomes quite long.

The constantly declining $\dot M$ of scenario B (where the initial
donor is a $0.05 \msol$ brown dwarf) gives a different thermal
outcome, as the WD constantly cools (dashed line in upper panel) and
its luminosity declines (dashed line in bottom panel). There are only 25 CNe in 6 Gyr; indeed, the donor itself only has enough mass for about 50
CN events.

\subsection{Dwarf novae in quiescence}
\label{sec:dwarf}

Both of these scenarios have low enough
$\dot M$ that the accretion disk should be unstable, resulting in
dwarf nova outbursts \citep{shafter92} and long periods of
quiescence.   When the accretion disk is in outburst, the accretion luminosity
\be
	L_{\rm acc} &\approx& \frac{ G M \dot{M} }{ 2R } \nonumber \\
	&\approx& 0.004 L_\odot \lp \frac{M}{0.4 \msol} \rp \lp \frac{10^9 {\rm \ cm} }{R } \rp \nonumber \\
	&& \times \lp \frac{ \dot{M} }{10^{-11} \smpy } \rp 
\ee
is an order of magnitude larger than the luminosity in the WD envelope, because the gravitational specific energy, $GM/R$, is much larger than the thermal specific energy, $k T_c / \mu m_p$, which sets the scale of the envelope's thermal profile.  However, when the disk is in quiescence, the excess $L_{\rm acc}$ is quickly radiated away on the thermal timescale near the WD's photosphere, and the luminosity of the system is dominated by the WD's quiescent surface luminosity, $L_q$, which originates from deep beneath the photosphere \citep{piro05}.  

\begin{figure}
        \plotone{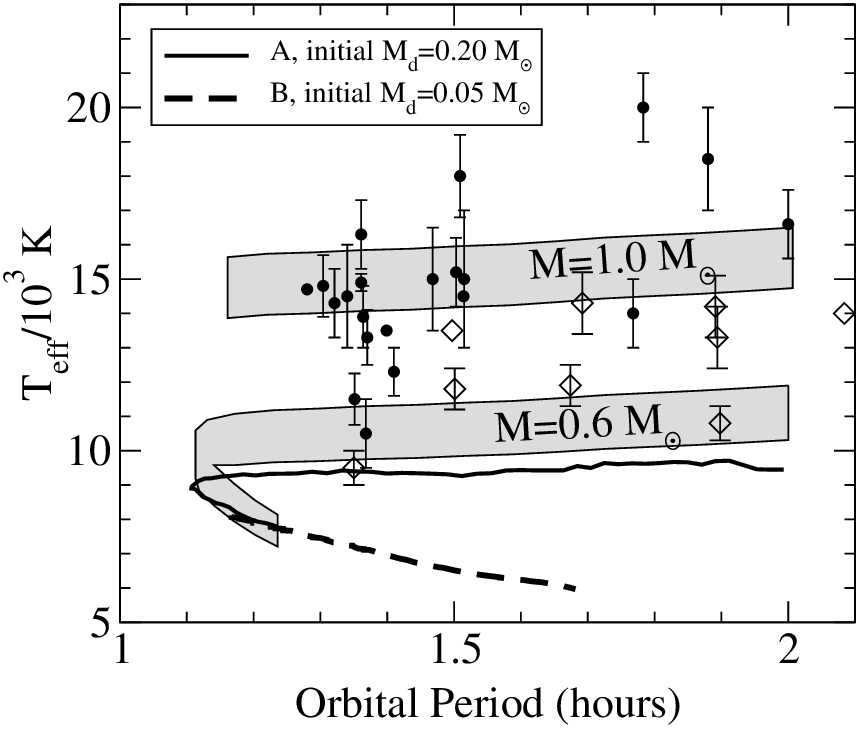}
       \caption{Adopted figure from \cite{tg09}. Solid (triangles)
         points are $T_{\rm eff}$ measurements from non-magnetic
         (magnetic) WDs in CVs below the period gap from \cite{tg09}.
         Shaded regions are the predicted effective temperatures found by
         \cite{tg09} for $0.6 \msol$ and $1.0 \msol$ WDs. The solid
         and dashed lines are our calculations for a
         $0.4 \msol$ He WD with a $0.2 \msol$ and $0.05 \msol$
         donor, respectively.  We use the minimum in the quiescent surface luminosity for calculating $T_{\rm eff}$. Figure \ref{fig:lqui} shows
that for 90\% of the CN cyle, the value of $T_{\rm eff}$ is no more than 15\% higher than the value we plot.}
       \label{fig:teffevol}
\end{figure}

Ever since \cite{sion99} discussed the value of $T_{\rm
eff}$ measurements of dwarf novae in quiescence, observers and
theorists have vigorously pursued this important diagnostic. Most
recently, \cite{tg09} summarized the observations and improved the
earlier theoretical work \citep{tb02,godon02,tb03,tb05,piro05}. In
Figure \ref{fig:teffevol}, we have added the $T_{\rm eff}$
evolutions for our two scenarios to the original figure from
\cite{tg09}. Some data are marginally consistent with the solid line
(scenario A), whereas no observed WDs are as cold as predicted from
our scenario B; however, it is unclear if the absence of these cold systems is physical or due to selection effects. It would be of interest to learn whether those cold
systems near the solid line have any other evidence for having low-mass accretors.

The continual discoveries of CVs in quiescence by the Sloan Digital
Sky Survey (e.g., \citealt{szkody09}; \citealt{gaen09}) and, in the near future, by
SkyMapper \citep{skymapper} will certainly provide new opportunities
to reveal accreting He WDs. In particular, \cite{gaen09}'s recent discovery of the expected ``pile-up'' of CVs at the $80-86$ min orbital period minimum has alleviated long-standing concerns regarding binary evolution.  Hence, as we noted in the introduction, $\approx 20\%$ of the CVs in \cite{gaen09}'s compilation should harbor a He WD.  The discovery and study of accreting WD
pulsators (see \citealt{muka07} for an updated list) may well be our best
hope, as \cite{arras06} noted that low-mass WDs have
colder effective temperatures for pulsation than higher mass WDs, and our $T_{\rm eff}$
calculations shown in Figure \ref{fig:teffevol} are slightly colder than the blue
edge calculated by \cite{arras06} for low gravity (e.g., low-mass
WDs). Thus, a prevalence of pulsators at low $T_{\rm eff}$ may be an
indicator of He WDs. Certainly more work is needed to make this
connection clear, but the rapid increase in the discovery of such pulsators is bound to
reveal a few new systems worthy of intensive study.


\section{Physics of classical nova outbursts}
\label{sec:indiv}

In the previous sections, we studied the secular evolution of the WD.  We now turn our attention to details of the individual nova outbursts, focusing on the evolution of the convective burning phase, the composition of the nova ejecta, and the WD's post-nova appearance as a supersoft x-ray source.


\subsection{Evolution of the convective envelope and maximum temperatures}
\label{sec:conv}

For most of the accretion phase of the nova cycle, the thermal profile is set by ``compressional'' heating (see \S \ref{sec:constMdot}).  However, once the base of the envelope becomes dense and hot enough, the energy generation rate from nuclear burning becomes large enough that radiative diffusion can no longer effectively transport the luminosity.  The radiative envelope is then transformed into a convective zone whose entropy is increased by continued nuclear burning until mass is lost via a radiatively-driven wind, and the nova outburst is observed.  We now give a brief overview of the relevant physics during this stage of the CN evolution; for a more complete explanation of similar physics in He-burning convective envelopes, see \S 3 of \cite{sb09b}.

\begin{figure}
	\plotone{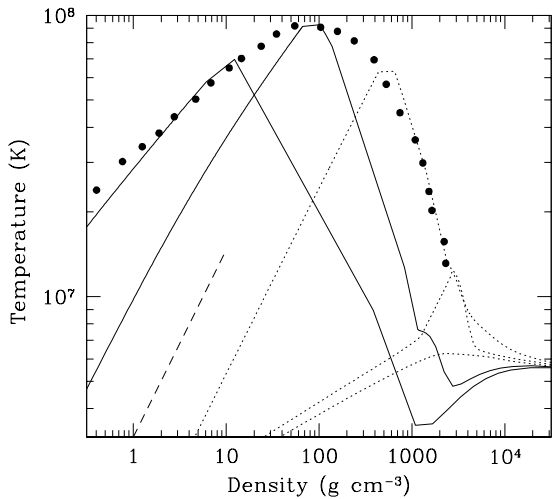}
	\caption{Density and temperature evolution during a CN on a $0.4 \msol$ He WD.
The dots display the evolution in time (from right to left) of the density and temperature at the base of the convective zone during the outburst. The uppermost dotted line is the
profile when the convective zone reaches the surface of our numerical grid. The dotted lines beneath
that are for times 5900 yr and $4.5\times 10^6$ yr prior. The solid lines show the profile at times 0.04 (when the base temperature reaches its maximum) and 2.1 yr after convection reaches the surface. The dashed line is the adiabatic slope, $\left. \partial \ln T / \partial \ln \rho \right|_s$, which is, as expected, nearly parallel to the convective region.  The radial expansion of the outermost shell reduces the pressure at the base
of the envelope, allowing for the underlying helium to adiabatically expand and cool (as is clear from the two solid lines at densities above $10^3 \ {\rm g \ cm^{-3}}$.}
	\label{fig:rhotzone}
\end{figure}

Figure \ref{fig:rhotzone} shows an example of the evolution of a convective envelope as it is heated by nuclear burning, numerically calculated for a $0.4 \msol$ WD that accreted mass at a rate of $10^{-11} \smpy$ after it has reached its equilibrium core temperature of $5.5\E{6}$ K.  Dotted lines show temperature-density profiles of the envelope at $4.5\E{6}$, $5900$, and $0$ yr before the convective zone reaches the top of our numerical grid, and solid lines show profiles $0.04$ and $2.1$ yr after this time, with more evolved envelopes at lower densities.  The dashed line is the adiabatic slope for an ideal gas, showing that the convective envelopes are indeed isentropic.

During the initial heating of the convective zone, the envelope is geometrically thin, with a scale height $h=P_b/ \rho_b g $ much less than the WD radius $R$, and so the pressure at the base remains nearly constant: $P_b \approx G M M_{\rm env} / 4 \pi R^4$.  Thus, as entropy is injected into the envelope during this phase, the temperature at the base increases and the density decreases while maintaining a constant pressure.  Bullets in Figure \ref{fig:rhotzone} demarcate the base of the convective zone, which, as predicted, initially becomes less dense and hotter at constant pressure as the entropy increases.

Further entropy injection causes the scale height to grow until it becomes a significant fraction of the WD's radius, and the envelope transitions to being geometrically thick.  This leads to radial expansion of the envelope and a subsequent decrease in the base temperature.  Thus, there is a maximum base temperature for the convective envelope, which depends only on $M$, $M_{\rm env}$, and the envelope's composition \citep{sb09b}.  For the model shown in Figure \ref{fig:rhotzone}, the maximum temperature is $\approx 10^8$ K.

\begin{figure}
	\plotone{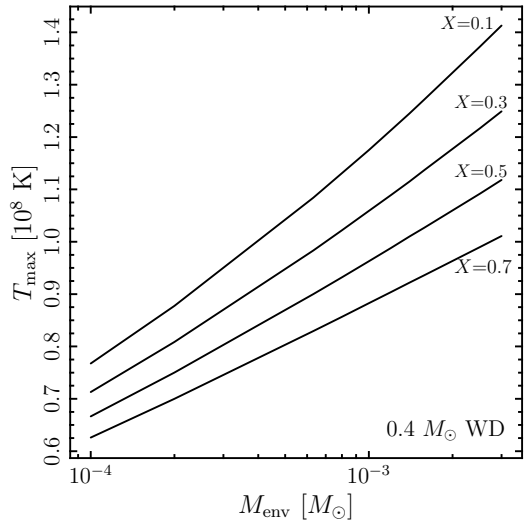}
	\caption{Maximum base temperatures vs. envelope mass for
          convective envelopes on a $0.4 \msol$ He WD.  The curves
          are for different values of the hydrogen mass fraction, $X$,
          in the convective envelope, which depend on the amount of core-envelope
          mixing.}
	\label{fig:tmax}
\end{figure}

Figure \ref{fig:tmax} shows analytic calculations of the maximum base temperature as a function of the
convective envelope's mass for a $0.4 \msol$ He core.  The four lines are calculated with different hydrogen mass fractions, $X$, in order to parametrize the amount of core-envelope mixing.  Increasing the helium abundance and thus the mean atomic weight $\mu$ decreases the number
density of particles, and so $T$ is higher for a given pressure.  Doubling $X$ decreases the maximum base temperature by $\approx 15\%$.  Note that the
temperatures for these models never reach values high enough to 
ignite the underlying He WD. Such an outcome would be exciting as it would likely lead to the birth of a
He-burning star in a tight orbit. 


\subsection{Composition of the nova ejecta}
\label{sec:comp}

Since the maximum  base temperature in the convective envelope is not very high ($\approx 10^8 $ K), the main channels for nuclear burning during the outburst are the $p$-$p$ chain and CNO cycle.  While CNe on C/O WDs can dredge up large amounts of $^{12}$C and $^{16}$O nuclei from the underlying WD core, the only source of CNO nuclei for CNe on He WDs is from the accreted material and any trace CNO left in the He core.  As a result, the ejecta has a metallicity, $Z$, close to the solar value, with an excess of H and He and a deficiency of $^{12}$C and $^{16}$O as compared to CNe on C/O WDs.

\begin{table}
  \begin{center}
    \caption{CN Ejecta Mass Fractions on He and C/O WDs}
\begin{tabular}{ |c| c| c| c|}
  \hline
  Element & Solar & Ejecta (He WD) & Ejecta (C/O WD) \\
  \hline
  \hline
H & 0.70 & 0.62 & 0.59 \\
\hline
He & 0.28 & 0.36 & 0.30\\
\hline
$Z$ & 0.017 & 0.021  & 0.11\\
\hline
$^{12}$C & $3.9\E{-3}$ & $0.27\E{-3}$  & $2.3\E{-3}$\\
\hline
$^{13}$C & $0.43\E{-4}$ & $0.87\E{-4}$ & $7.7\E{-4}$\\
\hline
$^{14}$N & $0.10\E{-2}$ & $0.87\E{-2}$ & $5.7\E{-2}$\\
\hline
$^{15}$N & $3.6\E{-6}$ & $0.28\E{-6}$ & $2.4\E{-6}$\\
\hline
$^{16}$O & $0.94\E{-2}$ & $1.0\E{-2}$ & $4.8\E{-2}$\\
\hline
$^{17}$O & $3.5\E{-6}$ &  $1.4\E{-3}$ & $3.0\E{-3}$\\
\hline
          \end{tabular}
  \label{tab:abun}
    \end{center}
\end{table}

Table \ref{tab:abun} shows the mass fractions for solar composition and for the numerically calculated ejecta of CNe on a $0.4 \msol$ He WD accreting at $\mdot = 10^{-11} \smpy$ and a $0.65 \msol$ C/O WD accreting at $\mdot = 10^{-9} \smpy$, computed after many nova cycles.  Both calculations yield ejecta composed of $\sim 10\%$ core material and $\sim 90\%$ accreted material.  This mixing results from the initiation of convection below the original core-envelope interface due to diffusion during the accretion phase.  However, because the cores are composed of different material, $Z$ is essentially solar for CN ejecta on He cores with a CNO breakdown that roughly reflects the abundances of the equilibrium CNO cycle, whereas the ejecta from the C/O WD CN has a metallicity that is 5 times larger than solar with overabundances of all of the CNO elements.  Note that the specific isotopic ratios depend strongly on the reaction rates that are used and should thus be regarded as a rough guide.  However, the abundance trends of H, He, and metallicity are a robust result.

Classical novae whose ejecta exhibit near-solar metallicities, and enhanced He and CNO-processed abundances can thus help to identify these systems.  Of the currently small number of $\sim 25$ CNe with measured ejecta abundances (for compilations, see \citealt{gehrz98,hk06}), none definitively show these observational signatures.  However, given the fact that there should be far fewer CNe from He WDs as compared to C/O WDs (see \S \ref{sec:conclusions} for an estimate of the rates), this is not surprising.


\subsection{Post-nova supersoft phase}
\label{sec:supersoft}

During a nova outburst, mass is ejected from the convectively mixed
envelope (containing both accreted and core material) via a
radiatively-driven wind.  This mass loss continues until the envelope
re-establishes a hydrostatic solution, with a luminosity close to
the red giant core mass-luminosity relation of \cite{pac70}.  The
H-rich envelope undergoes quasi-steady nuclear burning with a nearly
constant luminosity, decreasing in mass as H is converted to He until
the remaining envelope becomes too small to maintain a steady-state
burning solution
\citep{tt98,sh05}. For near-Eddington luminosity and a photospheric
radius roughly equal to the radius of a $0.4 \msol$ core, the effective temperature yields a bright UV or supersoft x-ray source \citep{kh97} that
lasts for as long as the envelope burns stably.

\begin{figure}
\plotone{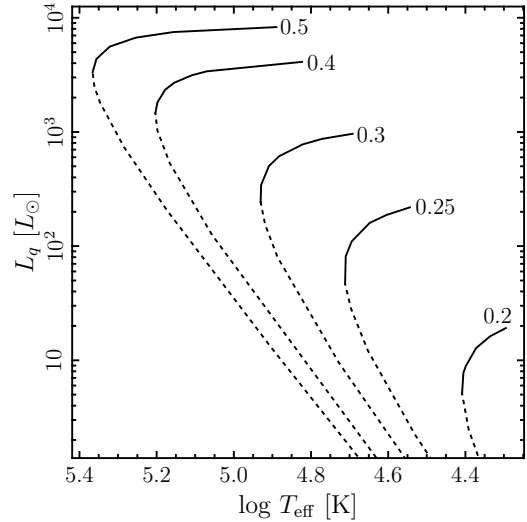} \caption{Evolution in the H-R diagram during the quasi-steady burning phase following a nova outburst. Each solid line shows the relation between $L$ and $T_{\rm eff}$ for a stably burning model for He core masses of 0.2, 0.25, 0.3, 0.4, and $0.5 \msol$.  Solid lines are truncated on the low $T_{\rm eff}$ side where the radius of the photosphere equals the Roche radius assuming a mass ratio of $1/2$ and $\porb = 4$ hr.  The dashed extensions are thermally unstable branches that are not physically realized.}
    \label{fig:lteff}
 \end{figure}

Specifying the core mass, envelope composition, and luminosity, and
demanding hydrostatic equilibrium and continuity of the radius and
pressure at the core-envelope interface uniquely determine $M_{\rm env}$ in this quasi-steady burning phase (see, e.g., Figs. 8
and 9 of \citealt{sh05} and Fig. 2 of \citealt{nom07}).  As the
envelope evolves, the luminosity remains nearly constant, decreasing
slowly while $T_{\rm eff}$ increases and the envelope mass decreases
until a minimum envelope mass is reached, which coincides with the
maximum $T_{\rm eff}$.

This initial phase of evolution is shown as
solid lines in the H-R diagram in Figure \ref{fig:lteff} for He WD
masses of 0.2, 0.25, 0.3, 0.4, and $0.5 \msol$.  The evolution
for these solutions begins to the right of the figure at the point where the photosphere is just inside the He WD's Roche radius assuming $M_d/M=1/2$ and $\porb=4$ hr, which we take to be an estimate of the end of the mass-loss phase.  The WD evolves towards the maximum $T_{\rm eff}$ and minimum $M_{\rm env}$ at the left.  The equation of state \citep{scvh95,ts00,rn02}, opacity \citep{ir93,ir96,ferg05,cass07}, electron
screening \citep{grab73,aj78,itoh79}, nuclear burning network
\citep{cf88,angu99,tim99}, and neutrino cooling \citep{itoh96} for these models
are calculated with the MESA code package\footnote{http://mesa.sourceforge.net/} (Paxton et al., in prep.).  The mass composition of the envelopes is 90\% solar composition (as defined by \citealt{lodd03}), which is accreted from the donor star, and 10\% He core material due to convective mixing during the thermonuclear runaway, chosen to roughly match the ejecta abundances in Table \ref{tab:abun}.  The CNO elements are assumed to be in chemical equilibrium as determined by their CNO-cycle burning rates, as their reaction timescales are shorter than the envelope's evolutionary timescale during this post-nova phase.  Table \ref{tab:SSS} shows various properties of the envelope at the maximum $T_{\rm eff}$ and minimum $M_{\rm env}$ for several core masses, where $\tau_{\rm SSS}$ is defined below, as well as the envelope mass at Roche lobe filling, $M_{\rm env,RL}$.  The maximum $T_{\rm eff}$ ranges from several to tens of eV.

\begin{table*}
  \begin{center}
    \caption{Post-Nova Conditions at Maximum $T_{\rm eff}$ / Minimum $M_{\rm env}$}
    \begin{tabular}{|c|c|c|c|c|c|}
      \hline
      $M$ & $M_{\rm env,RL}$ & $M_{\rm env}$ & $L_q$ & $T_{\rm eff}$ &
      $\tau_{\rm SSS}$\\
      ($M_\odot$) & ($10^{-3} \ M_\odot$) & ($10^{-3} \ M_\odot$) & ($L_\odot$) &
      ($10^5 {\rm \ K}$)  & (${\rm yr}$)\\
          \hline
          \hline
          0.2 & $3.1$ & $2.8$ & $5.1$ & $0.26$ &
          $3.7\E{7}$\\
          \hline
          0.25 & $ 1.1 $ & $0.92$ & $46$ & $0.51$ &
          $1.3\E{6}$  \\
          \hline
          0.3 & $ 0.52 $ & $0.37$ & $2.4\E{2}$ & $0.85$ &
          $1.0\E{5}$  \\
          \hline
          0.4 & $ 0.20 $ & $0.10$ & $1.4\E{3}$ & $1.6 $ &
          $4.8\E{3}$ \\
          \hline
          0.5 & $ 0.11 $ & $0.044$ & $3.3\E{3}$ & $2.3 $ &
          $8.9\E{2}$  \\
          \hline
          \end{tabular}
  \label{tab:SSS}
    \end{center}
\end{table*}

\begin{figure}
	\plotone{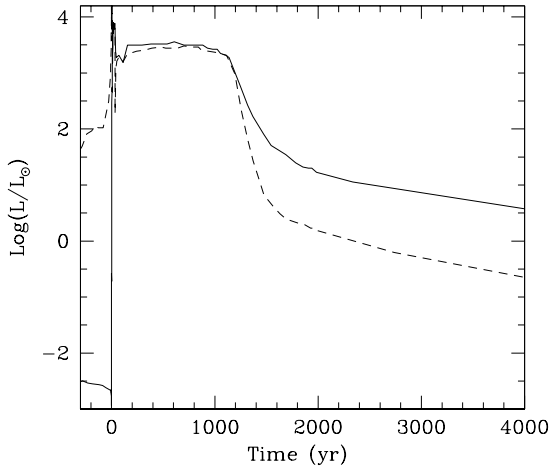}
	\caption{Integrated nuclear luminosity (dashed line) and quiescent surface luminosity, $L_q$, (solid line) versus time after a typical nova outburst on a $0.4 \msol$ He WD accreting at $10^{-11} \smpy$ after it has reached its equilibrium $T_c$.  The constant luminosity plateau lasts for $\approx 1000$ yr, after which the nuclear luminosity drops sharply and the WD cools.}
	\label{fig:lumevol}
\end{figure}

While solutions exist for lower luminosities than that at the maximum
$T_{\rm eff}$ and minimum $M_{\rm env}$, these solutions are not
physically relevant because they require the envelope mass to
\emph{increase} as time progresses \citep{sh05}.  Furthermore, a
stability analysis shows that these solutions are thermally unstable
\citep{pac83,nom07,sb07}.  These unrealized solutions are shown as
dashed lines in Figure \ref{fig:lteff}.  In reality, once the minimum
envelope mass has been reached, nuclear burning declines, the
envelope mass remains constant, and the WD evolves to the
WD cooling track until accretion leads to the next nova outburst.
Figure \ref{fig:lumevol} shows the time evolution of the integrated nuclear luminosity in the envelope and the quiescent surface luminosity from our numeric calculation after a typical nova outburst.  The quasi-constant luminosity phase lasts for $\sim 1000$ yr before burning declines.

We define a characteristic time the evolved post-nova spends near its
maximum $T_{\rm eff}$ as $ \tau_{\rm SSS} \equiv M_{\rm env} E_{\rm nuc} / L$,
where $ E_{\rm nuc} \approx 4.1\E{18} \ {\rm erg \ g^{-1}}$ is the energy per mass obtained by converting H to He in material with $X=0.68$.  As shown in Table \ref{tab:SSS}, these timescales are as long as 5000 yr for a $0.4 \msol$ core.  Given this long duration, it is of interest to look for long-lasting ($ \gtrsim 10 $ yr) supersoft sources with short orbital periods $<4$ hr.  Possible systems like this are RX J0537.7-7034, which has $\porb = 3.5$ hr and a dynamic mass estimate of $M=0.4-0.8 \msol$ \citep{oo93,greiner00}, and 1E 0035.4-7230 ($\porb = 4.1$ hr; \citealt{ori94,sch96,vrpb98}), whose closest non-LTE spectral fit yields $T_{\rm eff} \approx 3\E{5}$ K and $L \approx 2600 \lsol$ \citep{kph99}, which is consistent with a $0.5 \msol$ He WD near its maximum $T_{\rm eff}$.  Neither of these sources seems to fit the canonical explanation of supersoft x-ray sources as thermal timescale mass transfer systems \citep{kingsch01}.  Furthermore, \cite{pie05,pie07}'s growing catalog of M31 supersoft sources may also prove to be useful in this search for He WD post-nova supersoft x-ray sources.


\section{Conclusions}
\label{sec:conclusions}

Prompted by the lack of convincing detections of He WDs in CVs, we have studied their fate and thermal evolution in the hopes that observers may find new techniques with which to reveal or constrain this population.  We noted the existence of pre-CVs with He WDs in \S \ref{sec:masstransfer} and derived the mass-transfer evolution for two examples in \S \ref{sec:cvmdot}, under the assumption of angular momentum loss due to gravitational wave radiation.  In \S \ref{sec:example}, we described the thermal evolution of the core both analytically and numerically for several cases with constant $\mdot$, and in \S \ref{sec:fullscenario}, we described the reheating histories for the time-dependent CV evolutions of \S \ref{sec:cvmdot}.  In \S \ref{sec:indiv}, we focused on the details of the individual nova events, describing the convective burning phase (\S \ref{sec:conv}), the ejecta composition (\S \ref{sec:comp}), and the post-CN outburst quasi-steady burning phase (\S \ref{sec:supersoft}).

In \S \ref{sec:fullscenario}, we addressed the issue of the discovery of these systems while in quiescence.  We now turn to the question of their detection via their nova outbursts and post-nova burning phases.  From population synthesis predictions \citep{dek92,pol96,hnr01}, $\approx 20\%$ of CVs below the period gap harbor a He WD accretor.  If, for simplicity, we consider only CNe on $0.4$ and $0.6 \msol $ WDs, whose ignition masses differ by a factor of $\approx 2$ for a given $\porb$ \citep{tb05,yaron05}, we find that the rate of CNe from He WDs is $\approx 10\%$ of all CNe with orbital periods below the period gap.  However, for some novae, and especially extragalactic novae, the orbital period is much harder to measure, and so we cannot always divide the nova population into systems above and below the period gap.  From theory and observations \citep{warn02,tb05}, CNe below the period gap make up $10- 25\%$ of all CNe, and thus we expect He WD accretors in $1-3\%$ of all CNe with no $\porb$ information.  Note that this rough estimate does not take into account the selection effects that arise from the different peak luminosities and outburst timescales, which depend on the CV parameters.

As we showed in \S \ref{sec:supersoft}, the He WD remains bright in UV and supersoft x-rays during the post-nova burning phase.  Due to the similar physics at play (i.e., H-burning on low-mass degenerate cores), these post-novae have similar observed characteristics to low-mass post-asymptotic giant branch (AGB) and post-early AGB stars.  From \cite{bac06} and references therein, the post-AGB birth rate is $\sim 2$ per yr in a $10^{11} \msol$ E/S0 galaxy.  Extragalactic studies of CNe find a total rate of $20 \pm 10$ per yr in a galaxy of the same size \citep{ws04}, of which, from the calculation above, $ 0.2-0.6$ of these will be novae from He WDs.  Thus, there should be $ 10-30\%$ as many post-novae He WDs as post-AGB stars.

As these post-novae cool, their evolution takes them towards the pulsational instability strip.  \cite{arras06} examined the effect of envelope composition on pulsating WDs, and found that above a helium mass fraction of $\approx 0.3$, a HeII convection zone is present that results in a second instability strip, with a hotter blue edge than the canonical H/HeI strip.  If the post-nova envelope has a helium mass fraction of $ \approx 0.4$ due to core-envelope mixing, \cite{arras06} predict the blue edge of the resulting instability strip to be at $\approx 1.8\E{4}$ K, far hotter than the standard empirically-derived strip.


\acknowledgments

We thank Dean Townsley for a thorough review of a preliminary draft, Dina Prialnik and Attay Kovetz for use of their code, Boris G\"{a}nsicke, Tom Marsh, and Marina Orio for helpful discussions, and the referee for constructive comments.  This work was supported by the National Science Foundation under grants PHY 05-51164 and AST 07-07633.


\appendix
\label{sec:appendix}

Given the long accumulation timescales between novae on low-mass He WDs, we provide here an analytic estimate of the importance of chemical diffusion from the solar composition envelope into the He core.  If the diffusion timescale, $\tau_d$, is of order or less than the accumulation timescale, chemical diffusion will be non-negligible.  In the limit that $\tau_d$ is much shorter than the accumulation timescale, the composition will be in a diffusive equilibrium governed by a balance of the gravitational and electric forces.  For a degenerate He gas, the electric field is $e\vec{E} = -2 m_p \vec{g}$, where $\vec{g}$ is the gravitational acceleration, and so the net force on a hydrogen nucleus is $-m_p \vec{g}$.  In diffusive equilibrium, balancing the forces yields an exponentially decreasing hydrogen tail going into the He core with a length scale roughly equal to the pressure scale height.  Heavier elements like He, $^{12}$C, $^{14}$N, and $^{16}$O, on the other hand, experience no net force in the degenerate core because they have the same charge-to-mass ratio as helium, and so they perform a random walk inwards and will have a constant mass fraction in the core when in diffusive equilibrium.

The timescale governing diffusion over a scale height is $\tau_d
\sim H^2/D$, where $H=P/\rho g$ is the pressure scale height, and $D$
is the diffusion coefficient.  From  \cite{ai80}, the
diffusion coefficient for ions in a perfect gas of background plasma
is
\be
D = \frac{3 (2kT)^{5/2}}{16 n_1 (\pi m_1)^{1/2} Z_1^2 Z_i^2 e^4
  \Lambda_i} \approx 1 \frac{\rm cm^2}{\rm s} \frac{ T_7^{5/2} A_1^{1/2} }{ \rho_3
  Z_1^2 Z_i^2 \Lambda_i } ,
\ee
where $\rho_3=\rho/10^3$ g cm$^{-3}$, $T_7$ is the temperature in
units of $10^7$ K, $Z$ and $A$ are the ion's charge and atomic weight,
the Coulomb logarithm is $\Lambda$, the 1-subscript refers to the
background ion, and the $i$-subscript refers to the diffusing ion
species.  For the case of a degenerate He core, the scale height
is $H = 6.2\E{6} {\rm \ cm \ } \rho_3^{2/3} (5 / g_7) $, where $g_7$
is the gravitational acceleration at the core-envelope interface in units of $10^7$ cm
s$^{-2}$.  For a $10^{-3} \msol$ envelope on a $0.4 \msol$
core, $g_7 \approx 5$.

The effects of electron degeneracy are accounted for in the Coulomb
logarithm.  For electron degenerate material,
\be
\Lambda_i = \ln \left[ 1+\frac{(kT)^3 \eta^{1/2} \exp (\eta) }{4\pi
    n_1 Z_1^3 Z_i^2 e^6 } \right] = \ln \left[ 1+ 14.3 \frac{T_7^3 \eta^{1/2} \exp (\eta) }{\rho_3
    Z_i^2 } \right] ,
\ee
where $\eta$ is the degeneracy parameter and the second line has been evaluated for a background of helium.  The Coulomb ratio of Coulomb to thermal energies, $\Gamma_i \equiv E_{\rm Coulomb}/E_{\rm thermal}$, for He-He coupling near the outer edge of the core is $ 0.57 \rho_3^{1/3}/T_7$, so the helium is roughly an ideal gas.  With these approximations, the diffusion timescale through a scale height
becomes
\be
\tau_d \approx 2.4\E{6} {\rm \ yr \ } \rho_3^{7/3} \lp \frac{5}{g_7}
\rp^2 \frac{ Z_i^2 \Lambda_i }{ T_7^{5/2} } .
\ee
For a $10^{-3} \msol$ envelope halfway through its accumulation phase, the base
temperature is $5\E{6}$ K, the density at the base is $3\E{3}$ g
cm$^{-3}$, and so the diffusion timescale for H into the He core
is $\tau_d \approx 3\E{8}$ yr, assuming $\eta=2$.  For carbon diffusing inwards, $\tau_d
\approx 10^9$ yr.  Thus, for accumulation timescales $< 3\E{8}$ yr, the composition profiles will not be in diffusive equilibrium.

\begin{figure}
  \plotone{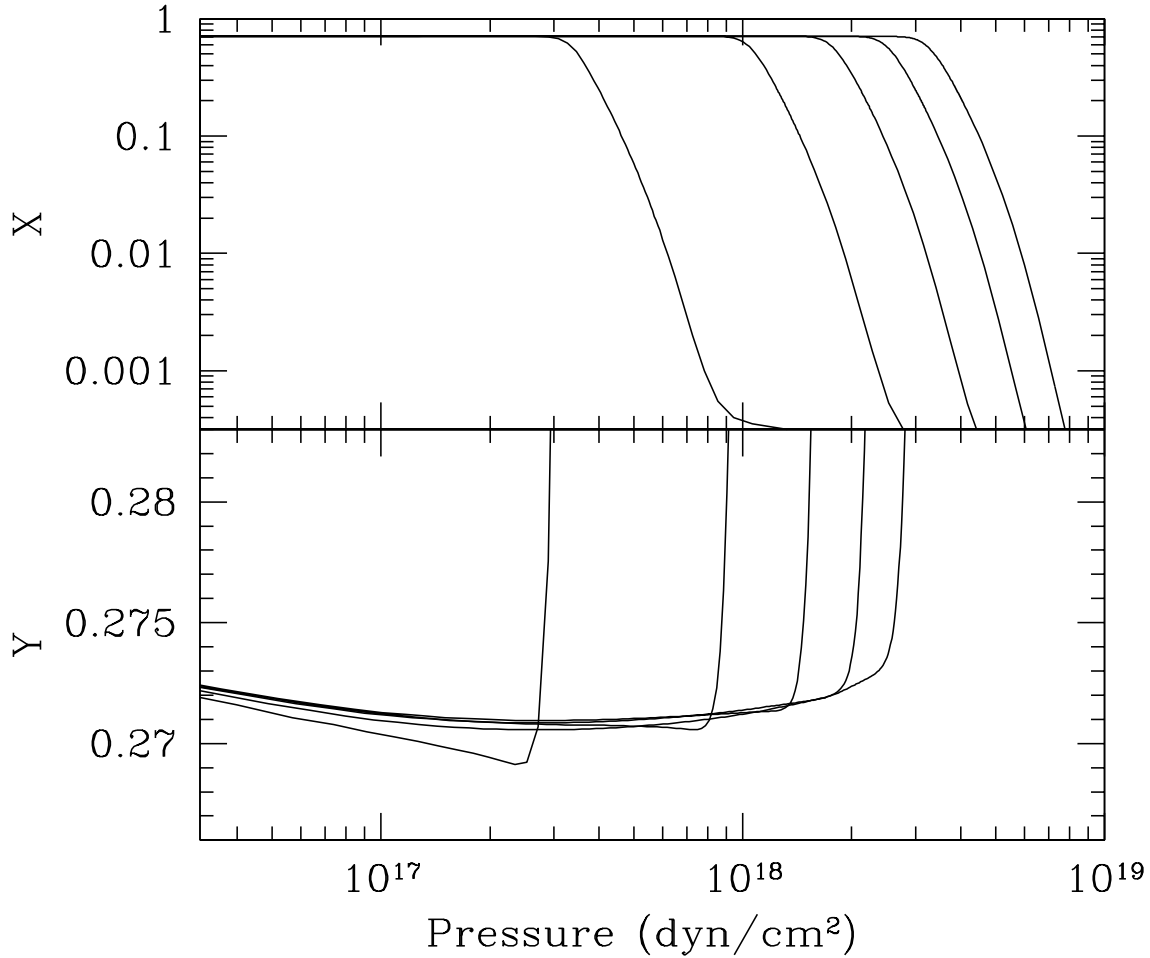}
  \caption{Hydrogen (top panel) and helium (bottom panel) 
mass fractions as a function of pressure for  different times
of an equilibrium CN cycle for $\dot{M}=10^{-11} \smpy$. From left to right, the curves are for times
of 10\%, 30\%, 50\%, 70\%, and 90\% of the $10^8$ yr CN cycle.  Note the different scales on the vertical axis.}
  \label{fig:Hdiff}
\end{figure}

Figure \ref{fig:Hdiff} shows the hydrogen (top panel) and helium (bottom panel) mass fractions as a function of pressure at different times during the $10^8$ yr accumulation phase of a $0.4 \msol $ He WD accreting at a rate of $\mdot = 10^{-11} \smpy$.  From lower to higher pressure, these profiles are at times of $10^7$, $3\E{7}$, $5\E{7}$, $7\E{7}$, and $9\E{7}$ yr after the previous CN.  Because the accumulation timescale is $10^8$ yr, neither the H nor heavier elements should be in diffusive equilibrium.  Indeed, as Figure \ref{fig:Hdiff} shows, the H is not yet in diffusive equilibrium; its exponentially-decreasing tail extends only $\approx 50\%$ of a pressure scale height inwards.



\end{document}